# Iron-silica interaction during reduction of precipitated silica-promoted iron oxides using *in situ* XRD and TEM

M.J. Coombes[ac*†], E.J. Olivier[a], E. Prestat[b], S. J. Haigh[b], E. du Plessis[c] and J.H. Neethling[a]

[a]Center for HRTEM, Nelson Mandela Metropolitan University, PO Box 77000, Port Elizabeth, South Africa

[b]School of Materials, The University of Manchester, Manchester, M13 9PL, UK

[c]R&T Analytics, Sasol Technology (PTY) Ltd., 1 Klasie Havenga Road, Sasolburg, South Africa

## Abstract

The effect of silica-promotion on the reduction of iron oxides in hydrogen was investigated using *in situ* X-ray diffraction and aberration-corrected transmission electron microscopy to understand the mechanism of reduction and the identity of the iron(II) silicate phase that has historically been designated as the cause of the iron-silica interaction in such materials. In the absence of a silica promoter the reduction of hematite to α-Fe proceeds via magnetite. Silica promoted amorphous iron oxide is reduced to α-Fe via stable magnetite and wüstite phases. During reduction of silica-promoted iron oxide, $Fe^0$ diffuses out of the amorphous silica-promoted iron oxide matrix upon reduction from $Fe^{2+}$ and coexists with an amorphous Fe-O-Si matrix. Certain portions of wüstite remain difficult to reduce to α-Fe owing to the formation of a protective silica-containing layer covering the remaining iron oxide regions. Given sufficient energy, this amorphous Fe-O-Si material forms ordered, crystalline fayalite.

**Keywords**

Silicon, Iron, Reduction, In situ, TEM

1. Introduction

Precipitated iron oxides, often prepared in the presence of various chemical and structural promoters, are common catalyst precursors used for the low-temperature Fischer-Tropsch (FT) process [1, 2]. Silica ($SiO_2$) is commonly employed as a structural promoter to stabilise the formation of small crystallites [3] and provides improved mechanical integrity, necessary for the catalyst to survive in the turbulent conditions of an FT slurry reactor [4]. However, the interaction between iron and silica inhibits the reduction of the iron oxide precursors to catalytically active phases and may result in a corresponding decrease in FT activity [5] [6] [7] [8] [9] [10] [11] [3]. This has long been linked to the formation of iron silicate phases during reduction.

Lund & Dumesic [7] first reported this iron-silica interaction in silica-supported magnetite ($Fe_3O_4$) samples which oxidised at high temperature to maghemite (ɣ-$Fe_2O_3$), with Mössbauer spectroscopy data suggesting $Si^{4+}$ cations substituting for $Fe^{3+}$ in the magnetite lattice, thus hindering transformation to hematite. Further reports proposed that this silicon-substituted magnetite exists as a 0.5 nm shell around an unsubstituted magnetite core [12]. This is supported by other studies which suggest the formation of thin iron(II) silicate rafts between iron oxides and the silica [13] or as encapsulating shells [8]. This Si-substitution has the effect of reducing catalytic activity, which is attributed to the iron cations in this region becoming more electron deficient and coordinatively more saturated with oxygen anions [14].

Dlamini *et al.* [10] demonstrated that the stage during synthesis which the $SiO_2$ is added influences its incorporation within the catalyst. When silica was added during or immediately after precipitation, the resulting iron oxide crystallites are small (approximately 3 nm) and exhibited the strongest iron-silica interactions during reduction. Zhang *et al.* reported on the synthesis, reduction and FT activity of two co-precipitated silica-promoted iron catalysts [11]. The presence of silica resulted in the formation of fayalite crystallites ($Fe_2SiO_4$) during FT synthesis, which hindered the formation of FT-active phases. Both Dlamini *et al.* and Zhang *et al.* observed iron species in the Mössbauer spectra which were ascribed to $Fe^{2+}$ associated with iron(II) silicates [10] [11].

An *in situ* X-ray microscopy study by Smit *et al.* [15] probed the composition of a silica-promoted iron FT catalyst in its oxide, reduced and FT-active states. Complex phase changes were observed, but the presence of an iron(II) silicate phase was reported in the reduced and FT-active sample using X-ray absorption spectroscopy. No iron(II) silicate phase was observed prior to reduction. Most recently, Suo *et al.* [3] characterised a series of binary iron-silica catalysts prepared by a co-precipitation method. They observed the formation of Fe-O-Si structures in the iron oxides which further transform into fayalite ($Fe_2SiO_4$) during FT synthesis.

Although iron-silica interactions during reduction have been widely studied, the characterisation methods employed to date are inherently bulk techniques, with an insufficient spatial resolution to accurately identify the location and mechanism of iron silicate phase formation in the reduction products. In the present work, we compare the morphology, elemental distribution/composition and phase composition of two co-precipitated iron oxide catalyst precursors after different temperatures of reduction in hydrogen. One of the iron oxide catalyst precursors was silica promoted and the other an unpromoted iron oxide catalyst precursor.

This study aims to determine the influence of the silica on the structural morphology of the promoted iron oxide catalyst precursor during reduction in hydrogen and to identify the location and type of possible iron silicate phases that may form, providing insight into the mechanism. Characterization was done using *in situ* powder X-ray diffraction (XRD) and transmission electron microscopy (TEM) which included electron energy loss spectroscopy (EELS) spectrum imaging and selected area electron diffraction (SAED) analysis. The findings are further supported by *ex situ* TEM analyses of the materials after reduction.

**2. Experimental**

2.1 Synthesis

The iron oxide samples were prepared using a continuous, co-precipitation technique [10, 11]. Briefly, precipitation was achieved by the dropwise addition of separate solutions of $Fe(NO_3)_3 \cdot 9H_2O$ and $Na_2SiO_3 \cdot 5H_2O$ at 80°C to approximately 75 mL of deionized water. The temperature of the water was maintained at 80°C and the pH was kept constant at 8.0 through the drop-wise addition of an $NH_4OH$ solution when necessary. After precipitation, the precipitate was immediately washed and filtered. The filter cakes were dried for 24 hrs at 90°C, ground to a fine powder and calcined in air at 400°C for 4 hrs. A promoted specimen (100/10 Fe/$SiO_2$) with 100 Fe: 10 $SiO_2$ by weight was compared with a similar unpromoted sample (similar synthesis but with no addition of silica). The specific loading for the silica promoted catalyst precursor was chosen since it allowed the findings to be industrially relevant and allow for detectability of Si during spectroscopic TEM analyses. Chemical promoters for precipitated FT catalysts (Cu and K) were consciously omitted from the synthesis to allow the Fe-O-Si interaction to be studied in isolation. It is assumed that the chemical promoters do not significantly influence the Fe-O-Si interaction.

The unpromoted (silica-free) iron oxide consists of irregular-shaped hematite (α-$Fe_2O_3$) crystallites, approximately 50 nm in diameter. The promoted iron oxide consists of small crystallites (2 – 3 nm) of silicon-containing amorphous iron oxide (am-Fe-Si-O). Characterisation data are presented in data-in-brief.

<u>2.2 Characterisation Techniques</u>

*2.2.1 In Situ Powder X-Ray Diffraction*

*In situ* powder, X-ray diffractograms were obtained using an Anton Paar XRK600 reaction chamber connected to a Malvern PANalytical diffractometer equipped with an Excellerator detector and a cobalt radiation source ($\lambda$ = 0.178897 nm). Samples were packed into a stainless-steel sample holder. Hydrogen gas with a flow-through configuration was introduced at 1.86 bara pressure at a flow rate of 50 mL.min$^{-1}$. A 2-hour scan was performed initially at 40°C, with a 20 min scan after every 20°C step increase, beginning at 80°C and ending at 360°C. The heating rate between scans was 10°C.min$^{-1}$. Four consecutive scans of 2 hours each were performed at a final temperature of 380°C. All scans were performed from $2\theta_{start}$ = 5° to $2\theta_{end}$ = 105°, with a step size = 0.017° $2\theta$. Crystalline phases present in the diffractograms were identified using Bruker AXS DIFFRAC.EVA software, while average crystallite sizes were determined using Rietveld refinement, the fundamental parameter, full pattern refinement approach and Topas4.1© software.

*2.2.2 Transmission Electron Microscopy – In Situ Reduced*

Aberration corrected TEM, STEM and EELS imaging was performed on an FEI Titan 80–200 ChemiSTEM equipped with a probe-side aberration corrector, an X-FEG electron source and a GIF Quantum EEL spectrometer. The experiments were performed using an acceleration voltage of 200 kV. *In situ* reduction was performed using a Protochips Atmopshere™ TEM Environmental Gas Cell holder. A small amount of sample was ultrasonicated in ethanol to disperse the iron oxide crystallites. A drop of this solution was placed on the silicon nitride window of the gas cell, which was subsequently assembled on the tip of the sample holder and checked for leaks before transferring to the TEM. Samples were reduced in high purity hydrogen gas (99.999 %) between 700 and 950 mbar absolute, while temperatures ranged from room temperature to 750°C. Regions of interest on the sample were consecutively studied at various temperatures to study changes in the iron oxides which occurred as a result of reduction in hydrogen. Each reduction step was done for 1 hr i.e. the cell was loaded with hydrogen at the target temperature and sealed for 1 hour. Thereafter it was evacuated to quench the reaction and all imaging and spectroscopic analysis performed. After imaging was complete, hydrogen was reintroduced at a new, higher temperature. Imaging and analysis of the samples were done in scanning transmission electron microscopy (STEM) mode using an angstrom sized probe. Probe current conditions (200 – 300 pA) were selected to optimize signal generation but, at the same time, minimize the risk of beam damage to the specimen. The convergence semi-angle of the probe was 21 mrad and the EELS acceptance semi-angle was 62 mrad. The EELS spectrum imaging was done using a 0.25 eV or 0.5 eV energy channel width for an energy range containing 2048 channels. The full-width at half maximum (FWHM) of the zero-loss peak was measured as approximately 1.3 eV. Selected area electron diffraction (SAED) analysis and hollow cone diffractive imaging were done in TEM mode using parallel illumination.

*2.2.3 Transmission Electron Microscopy – Ex Situ Reduced*

*Ex situ* TEM was performed using a double-aberration corrected JEOL JEM-ARM 200F operated at 200 kV and equipped with an Oxford Xmax80 EDS detector and Gatan GIF 965ERS with dual EELS capability. Reduced samples were prepared by placing a small amount of oxide powder on a glass slide on a metal susceptor, which was itself placed in a quartz reactor tube. The samples were reduced at 400°C (heating rate 10decC/min from room temperature) for 20 hours in ultrahigh purity hydrogen gas at a flow rate of 500 mL.min$^{-1}$. The reduced samples were immediately placed into ethanol and ultrasonicated to ensure they were well dispersed. A drop of dispersed solution was placed on an amorphous carbon film supported by a copper grid and immediately mounted on the sample holder and inserted into the microscope. The transfer to the TEM was performed as quickly as possible to minimise oxidation of the sample from exposure to atmospheric oxygen and moisture. Imaging and analysis of the samples were done in STEM mode using a sub-angstrom sized probe with a probe current between 68 pA and 281 pA. Probe current conditions were selected to optimize the beam current but, at the same time, minimize the risk of beam damage to the specimen. The convergence semi-angle of the probe used was fixed at 23 mrad with acceptance semi-angles of the GIF and dark-field detector being 84 mrad and 34 to 137 mrad respectively. The bright-field (BF) detector acceptance semi-angle was set at 0 to 12 mrad by using an illumination limiting aperture. The EELS spectrum imaging was done using a 0.25 eV or 0.5 eV energy channel width for an energy range containing 2048 channels. The FWHM of the zero-loss peak was measured as 1.5 eV. The SAED analysis was performed in TEM mode using parallel illumination.

**3. Results and Discussion**

## 3.1 XRD

Data obtained from an *in-situ* powder X-ray diffraction study (XRD) of the catalysts reduced in hydrogen are shown in Figure 1. Figure 1 (a) and (b) show X-ray diffractograms obtained after different temperatures of reduction for the unpromoted and promoted material, respectively. Figure 1 (c) and (d) is a diagrammatic representation of the evolution of phase composition for the unpromoted and promoted material with increasing temperatures of reduction, respectively. Figure 1 (e) and (f) shows the evolution of phase-specific average crystallite size for the unpromoted and promoted material with increasing temperatures of reduction, respectively.

We begin by discussing the interpretation of this data for the unpromoted sample. In the absence of silica ( Figure 1 (a), unpromoted), the reduction of the hematite ($\alpha$-$Fe_2O_3$) crystallites proceeds via the well-studied two-step pathway [16] [17] [18]. Firstly, the hematite is reduced to magnetite ($Fe_3O_4$), with the onset of the reduction occurring at 240°C. The relative phase abundance of magnetite achieves a maximum of 95 mass % at 300°C. Along with this, the magnetite average crystallite size increases to a maximum of 73 nm at 360 °C. At 320°C, a minor amount (less than 5 mass %) of wüstite (FeO) is formed, likely due to the reduction of magnetite. Wüstite is expected to undergo disproportionation at temperatures below 567°C [19], forming magnetite and iron. The absence of iron in the diffractogram at this temperature suggests that the wüstite is being stabilised by some factor, most likely a high water to hydrogen molar ratio within its local vicinity [20] [21] [22] [23] or a result of the slow kinetics which govern the disproportionation [24]. Metallic iron ($\alpha$-

Fe) formation is initiated at 340°C, although a small amount of wüstite remains present. The α-Fe shows a gradual increase in average crystallite size and abundance with all the iron oxides fully reduced to α-Fe at 380 °C with an average crystallite size of approximately 70 nm.

The $SiO_2$ promoted sample has a very different diffractogram from the promoted sample. Before heating the promoted sample occurs as amorphous silicon doped iron oxide (am-Fe-Si-O), with a diffractogram characteristic of two-line ferrihydrite [25] [26] [27], a highly-disordered iron(III) oxide, as evidenced by the two broad reflections at approximately 42 and 75° 2θ (corresponding to d-spacings of 2.5 and 1.5 Å respectively) seen in Figure 1 (b). Upon heating in the hydrogen atmosphere, there is no change in the reflection at 2.5 Å until 240°C, where the reflection grows slightly sharper. This is further exemplified at 260°C, while at 280°C, the magnetite {311} reflection is evident at this 2θ position (d-spacing). The reflection at 1.5 Å gradually shifts on heating to lower values of 2θ, indicating a larger d-spacing value. At 280°C it presents as a crystalline {440} reflection of magnetite. These changes in the short range order from 40°C to 280°C indicate a slow reduction of $Fe^{3+}$ cations in the silicon doped am-Fe-Si-O. $Fe^{3+}$ cations in the poorly ordered iron-oxygen clusters of am-Fe-Si-O reduce to $Fe^{2+}$ and the constituent atoms rearrange into a poorly ordered form of magnetite which slowly increases in crystallinity, up to 280°C, where the thermal energy is enough to arrange into a more crystalline structure. In previous work the reduction of unpromoted synthetic 2-line ferrihydrite was found to exhibit crystalline reflections of magnetite at the lower temperature of 250°C, suggesting that here the presence of silica is inhibiting magnetite formation [28].

Upon increasing the temperature of reduction from 280°C to 360°C, magnetite is seen to gradually reduce to predominantly wüstite. A small proportion (phase composition of 1 mass %) of α-Fe phase is detected at 360°C which shows the reduction of wüstite to metallic Fe. Within the same temperature region, the average crystallite size of magnetite is seen to moderately increase to a maximum value of 30 nm at 360°C with the wüstite stabilizing to an average crystallite size of between 26nm and 31nm at 360°C. Wüstite is seen to reduce to α-Fe starting at 380°C with the abundance of α-Fe increasing to 15 mass %. Continued reduction of the catalyst at 380°C for greater than 8 hours showed a significant proportion (57 mass %) of wüstite remaining in the material with both α-Fe and wüstite average crystallite sizes stabilising in the range of 26 nm to 28 nm.

It was also noted that in the silica-promoted sample, broad reflections at 2.5 and 1.5 Å ascribed to the am-Fe-Si-O phases are still present, even after reduction at 380°C for 8 hours indicating the persistence of some amorphous iron oxide even after such a lengthy reduction.

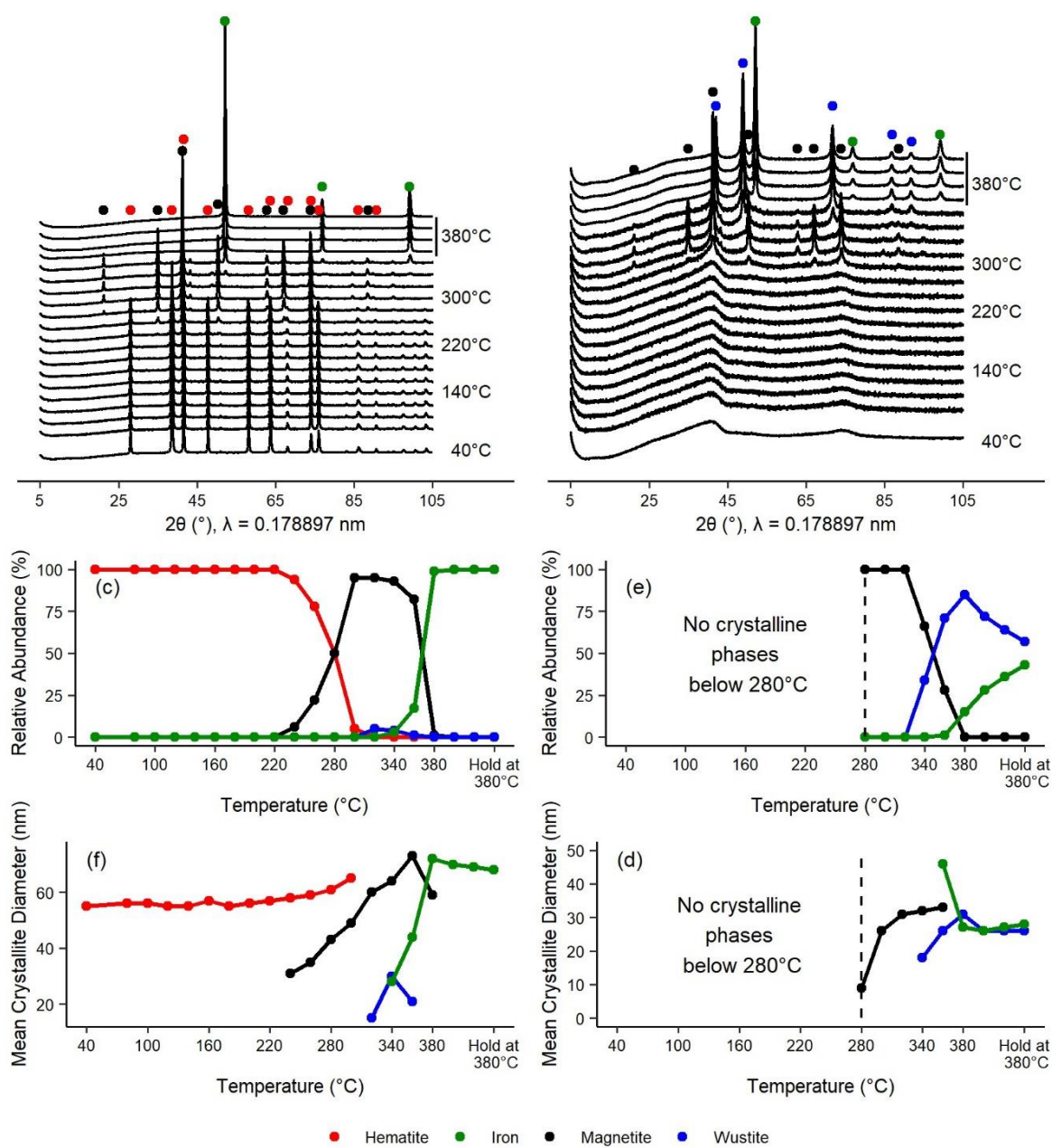

*Figure 1:* In situ *XRD patterns of (a) unpromoted, (b) promoted iron oxide samples. The relative abundance and crystallite size of the crystalline phases for (a) is shown in (c) and (e) respectively. The relative abundance and crystallite size of the crystalline phases for (b) is shown in (d) and (f) respectively.*

3.2 *In situ* TEM

The *in situ* TEM reduction of the unpromoted and promoted samples was performed to fundamentally understand the mechanism of reduction, at the spatial resolution offered by TEM.

**Unpromoted:** 100/0 Fe/SiO$_2$ was reduced in approximately 1 bara hydrogen atmosphere from room temperature to a temperature of 700°C. In Figure 2 the morphological evolution of a region of interest is shown after reduction temperatures up to 700°C using a series of annular dark field (ADF) STEM images acquired after the indicated reduction temperatures. The top part of Figure 3 shows selected area electron diffraction (SAED) patterns obtained from the region of interest after reduction at temperatures up to 700°C. The bottom part of Figure 3 shows electron diffraction spectra (normalised at maximum peak intensity) generated by rotationally averaging the SAED patterns shown in Figure 3 (top) about the centre point using a digital micrograph [29] plugin PASAD tools [30].

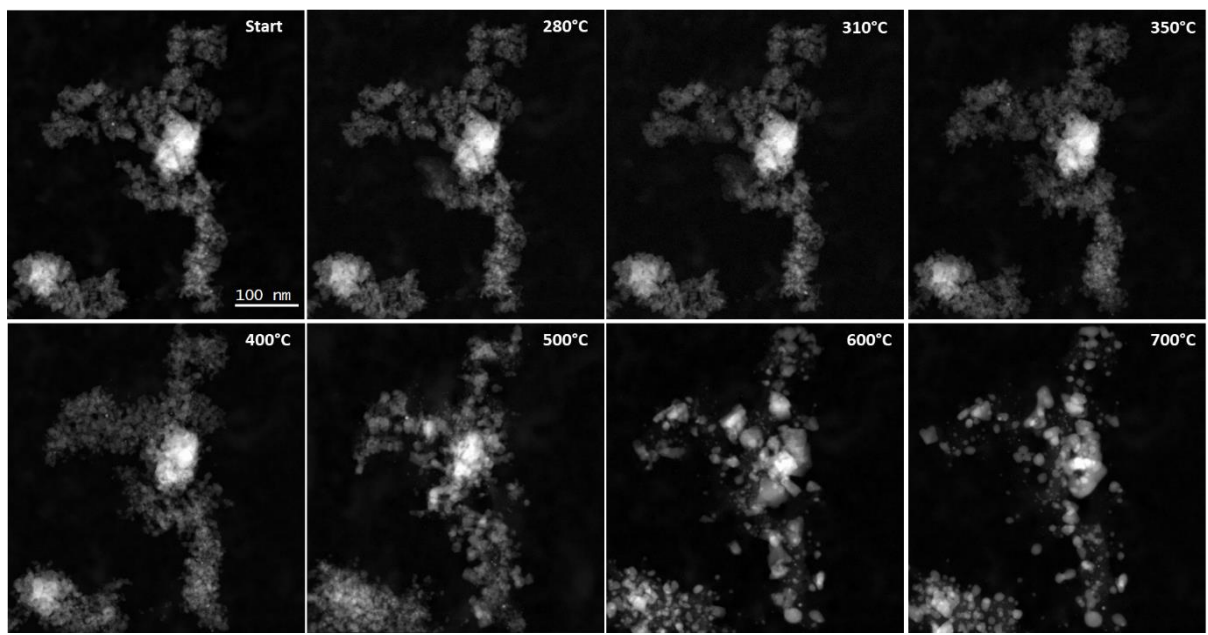

*Figure 2: A series of* in situ *ADF STEM images showing the morphology of a region of interest after reduction at the indicated temperatures for 1hr*

The first evidence of phase transformation owing to reduction was noted at 350°C (Figure 3) where the first occurrence of magnetite related electron diffraction peaks was observed. The observation is different from the *in situ* XRD data which suggests that the reduction of hematite to magnetite should have been largely complete at this temperature. The observed difference is attributed to the time at reduction temperature, potential underestimation of the temperature due to the presence of hydrogen in the TEM gas cell, differences in detection using a bulk technique (XRD) compared to a microscopic technique (TEM), as well as different configurations of the *in situ* reactors. The *in situ* XRD reactor has a flow through configuration with a gas flow rate of 50 mL/min while the *in situ* TEM cell is a closed, sealed system. Only slight morphological changes are observed in the area after reduction at 350°C as compared to reduction after 310°C (Figure 2). Further reduction at 400°C and 500°C show a gradual and complete transition from hematite to magnetite in the area (Figure 3) with significant changes in the morphology observed after reduction at 500°C (Figure 2). The mesoscale morphology of the area, however, remains largely the same, suggesting that reduction of hematite to magnetite is linked to combined sintering and reduction of hematite crystallites which are in proximity. With the further reduction of the region of interest at 700°C, only α-Fe phases are identified from the electron diffraction data (Figure 3) with additional sintering of crystallites observed. Nanoscale crystallites were also observed between larger crystallite phases and were identified as metallic iron nanoparticles using elemental mapping of a representative region using EELS spectrum imaging (shown in data-in-brief). The formation of small iron nanoparticles from comparatively large magnetite crystallites suggests a migration of iron atoms from the magnetite lattice upon reduction to the zero-valent state. This is discussed in more detail below.

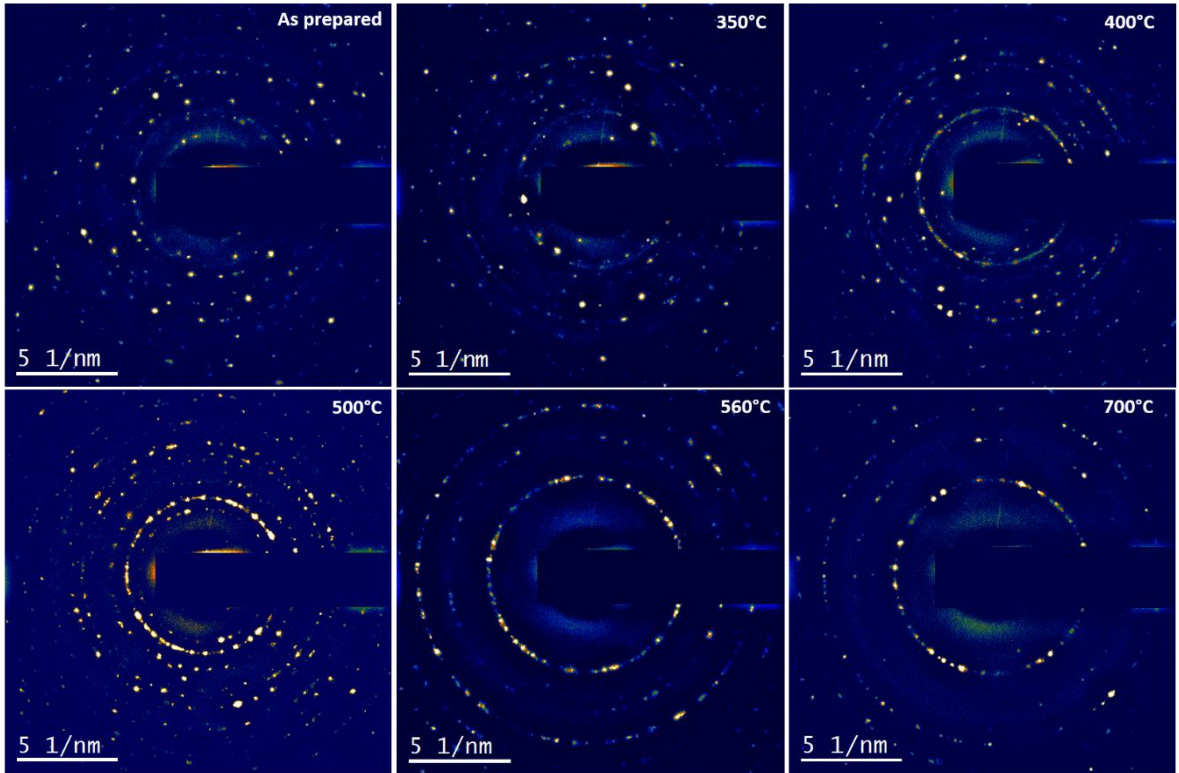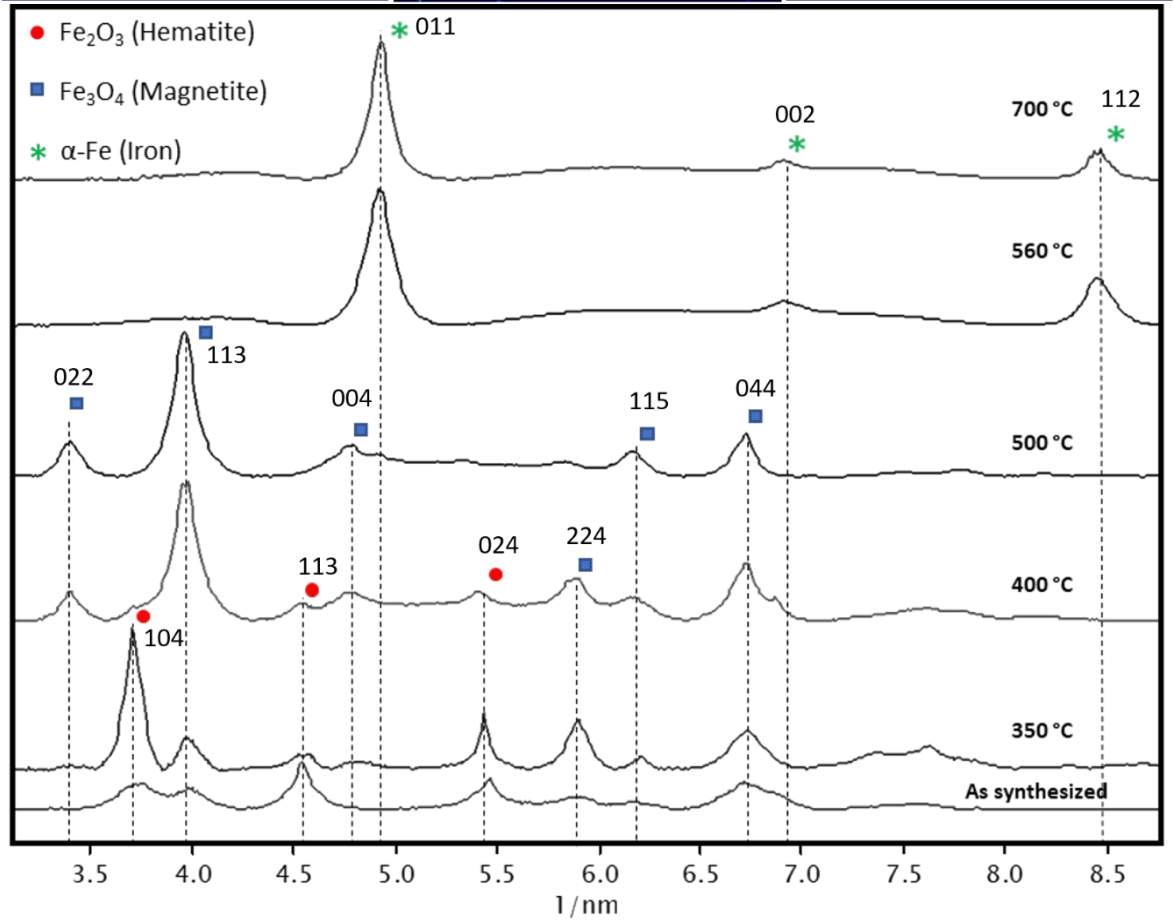

*Figure 3: Upper panel: SAED patterns obtained from the region of interest shown in Figure 2 after reduction at the indicated temperatures. Lower panel: electron diffraction spectra generated by rotational averaging of the SAED patterns shown above.*

**SiO$_2$ promoted**: The reduction of 100/10 Fe/SiO$_2$ in hydrogen was similarly studied using *in-situ* TEM.

Figure 4 shows a series of ADF STEM images representing the morphological evolution of a region of interest containing a silicon-substituted ferrihydrite after reduction at temperatures up to 750°C. Figure 5 (top) shows SAED patterns obtained from this region of interest after reduction at the indicated temperatures. Figure 5 (bottom) shows electron diffraction spectra generated by rotational averaging of the SAED patterns in Figure 5.

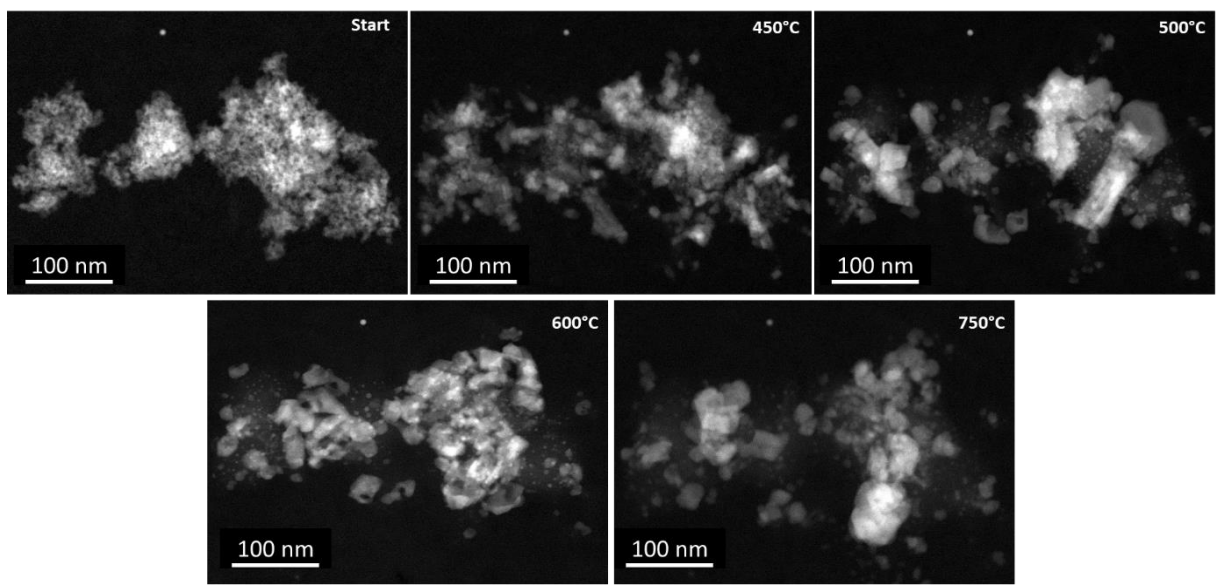

*Figure 4:* In situ *ADF STEM images showing the morphology of a region of interest of the promoted sample after reduction at the indicated temperatures.*

The as-prepared material (Figure 4) comprises of silicon-containing amorphous iron oxide. The diffraction rings seen in the electron diffraction patterns (Figure 5) for the as-prepared material show a combination of rings related to the silicon doped am-Fe-Si-O as well as the amorphous $SiN_x$ windows used to contain the gas for the *in situ* analysis. The first evidence of structural changes of the ferrihydrite was observed after reduction at 400°C with the presence of faint diffraction spots observed in the SAED pattern obtained from the region of interest (Figure 5). The first evidence of morphological changes (Figure 4) of the area was only observed after reduction at 450°C with evidence of sintering of am-Fe-Si-O regions observed. The crystalline phase formed during this transition was identified as magnetite by electron diffraction (Figure 5). The first evidence for the formation of wüstite in the area was only observed after reduction at 600°C, co-existing with magnetite. Wüstite was differentiated from the (004) line of magnetite by a slight shift from 4.7 1/nm for magnetite (004) to 4.6 1/nm for wüstite (002). This was further evidenced by the disappearance of the intense (022) and (113) peaks for magnetite but remaining presence of the 4.6 1/nm (002) peak for wüstite after reduction at 700°C. Further morphological changes of the region of interest occurred after reduction at 500°C and 600°C with a noticeable change in the mesoscale morphology from that of the original am-Fe-Si-O morphology. With the reduction at 700°C and 750°C, the emergence of diffraction peaks consistent with α-Fe, fayalite and wüstite formation was observed (Figure 5). Fayalite is the iron(II) silicate material that has long been attributed as the source of the iron-silica interaction which inhibits reduction to the catalytically active phase. The formation of magnetite, wüstite and α-Fe agrees with the *in-situ* XRD experiments. However, the formation of crystalline fayalite was likely not observed in the XRD data due to the maximum temperature of the *in situ* XRD being 380 °C.

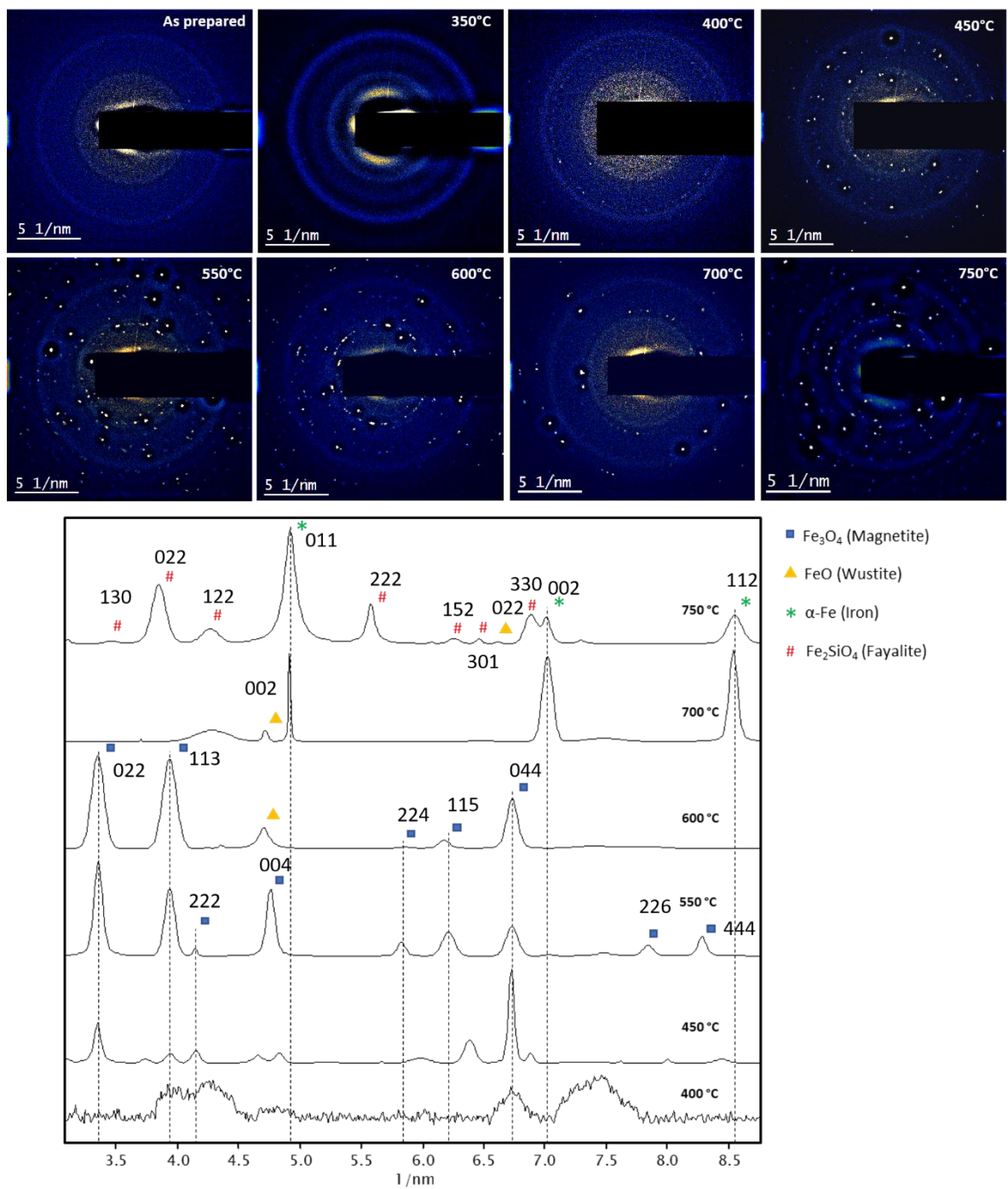

*Figure 5: Upper panel: SAED patterns obtained from the region of interest shown in Figure 4 after reduction at the indicated temperatures. Lower panel: electron diffraction spectra generated by rotational averaging of the SAED patterns shown above.*

The presence of nanometre-scale crystallites (nanoparticles) between larger sintered crystal phases was also observed in this case and evidence is provided in the data-in-brief.

The elemental distribution of Fe, O and Si, in the promoted sample as well as the Fe:O ratio, was determined using EELS spectrum imaging. EELS element quantification (with plural scattering removed) was done using a model-based quantification routine (excluding ELNES) as implemented in digital micrograph [29]. Figure 6 (left panel) shows the distribution of Fe, O and Si in the sample within the region shown in Figure 4 after reduction at 450 °C. Si exists in both the specimen and in the $SiN_x$ windows used to encapsulate the microreactor. To attempt to separate the two signals the Si to N ratio across the area was mapped by dividing the Si compositional maps with the N compositional maps. The Si to N ratio of the windows is expected to stay constant across the region (within error), thus, areas showing higher Si/N ratios can be attributed to Si present in the catalyst sample. Evidence of regions containing concentrated Si in the region of interest are shown in the Si/N map in Figure 6 (left). Also shown in Figure 6 (left) is a map of the compositional ratio of Fe to O in the area which identifies areas of Fe enrichment due to reduction. Figure 6 (right) is a representative diffraction pattern and false coloured hollow cone dark field diffractive TEM image of the area obtained after reduction at 450°C showing areas in the region of interest where the first magnetite (since only magnetite was identified from diffraction data) formation occurred. The hollow cone image was generated with a beam tilt of approximately 2.4 mrad using an objective aperture spanning an electron scattering angular range of 0.871 mrad (indicated by the red circle in Figure 7 (right panel). This enabled a hollow cone dark field image to be created using diffracted electrons originating from phases (magnetite determined from electron diffraction) with interplanar spacings between 0.174 nm (5.75 nm$^{-1}$) to 0.119 nm (8.4 nm$^{-1}$). The limits of this diffraction condition are indicated in Figure 6 (top-right) by the white circles. Areas of higher intensity in the hollow cone image indicate zones of magnetite formation.

From the hollow cone dark field image and Fe/O map, a qualitative correlation between sites of Fe enrichment and magnetite formation is observed. However, some regions of enriched Fe do not correlate with the presence of crystalline magnetite. In the absence of other phases identified at this temperature from electron diffraction, it is reasonable to attribute this discrepancy to the formation of poorly crystalline magnetite regions as proposed from observations of the XRD data.

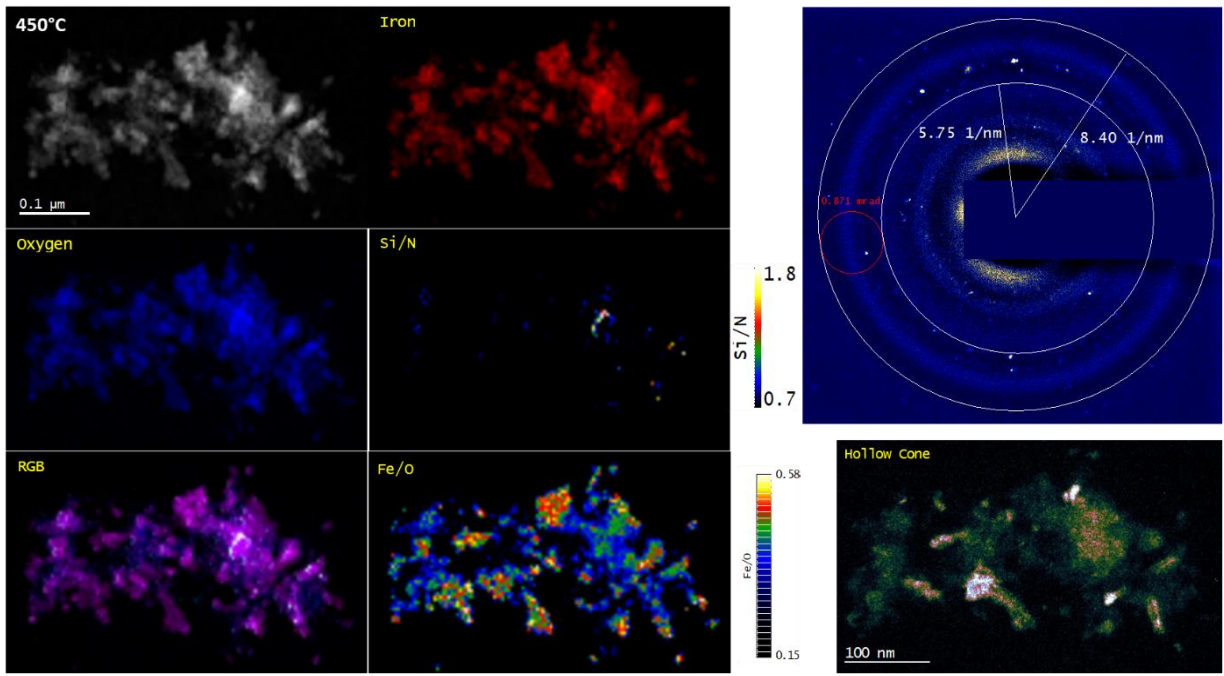

*Figure 6 (Left) Elemental distribution maps of Fe, O and Si of the region of interest generated from EELS spectrum imaging data. Included is a ration map of Si to N concentration in the area as well as Fe:O ratio. (Right) Representative diffraction pattern showing the region in diffraction space from which the hollow cone diffractive image was generated. Below is the generated hollow cone diffractive image showing the sites of magnetite formation.*

3.3 *Ex situ* TEM

**Unpromoted:** An *ex situ* TEM study of 100/0 Fe/SiO$_2$ after reduction at 400°C in hydrogen for 20 hours revealed the presence of large, crystalline particles similar to those observed after in situ reduction (Figure 4). A TEM micrograph representative of the general findings is shown in Figure 7 (a). The corresponding SAED pattern (Figure 7 (b)) contains several diffraction spots which identify the iron phase present as α-Fe. This is consistent with the *in situ* XRD results, which indicated complete reduction to α-Fe by 380°C. Evidence of a surface re-oxidation of the α-Fe crystallite to magnetite was also observed during analysis and discussed in the data-in-brief.

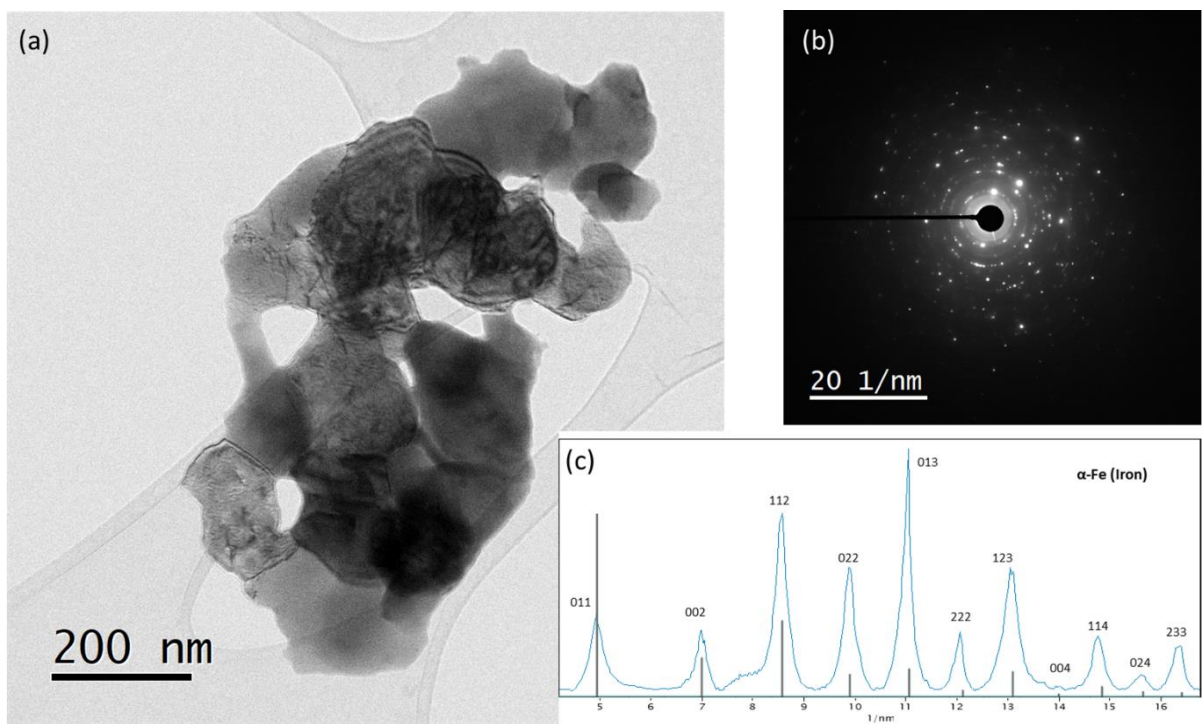

Figure 7: TEM analysis of unpromoted catalyst reduced ex-situ at 400°C for 20 hours in hydrogen. (a) BF TEM demonstrating the presence of large crystallites. (b) SAED pattern identifies α-Fe as the phase present. (c) Electron diffraction spectrum generated from (b) by rotational averaging identifying the crystallographic planes associated with the diffraction peaks.

**SiO$_2$ promoted:** TEM micrographs of the silica promoted Fe/SiO$_2$ after reduction at 400°C in hydrogen for 20 hours reveal a different particle morphology to the unpromoted catalyst. Figure 8 (a) shows an ADF STEM image of such a region. From the image, it is deduced that a mixture of crystalline phases is still present with some image characteristics consistent with an amorphous material present. A SAED pattern obtained from the red dotted circle shown in (a) is shown in Figure 8 (b) along with an electron diffraction spectrum shown in (c) generated by rotational averaging of the diffraction pattern in (b). The data indicates that the crystalline material is comprised of a mixture of magnetite, wüstite and α-Fe. The presence of a small amount of magnetite is likely a result of re-oxidation of α-Fe due to exposure to air during sample loading. Figure 8 (d) shows element distribution and composition maps for Fe, O and Si in the region generated from an EELS spectrum image obtained from the region. The formation of Fe enriched zones, consistent with α-Fe formation, is seen amongst iron oxide regions accompanied by segregation of Si to iron oxide regions. The concentration of the Si in the iron oxide regions vary between 1 and 13 at.%. A higher magnification EELS spectrum imaging analysis of an enriched Si area showed the presence of distributed silica networks between iron oxide crystallites. Elemental distribution maps of Fe, O and Si are shown along with a false colour overlay (RGB) of the signals obtained from a representative area in Figure 9. The finding is consistent with previous observations made from Mössbauer spectroscopy data inferring the presence of iron(II) silicates [10] [11].

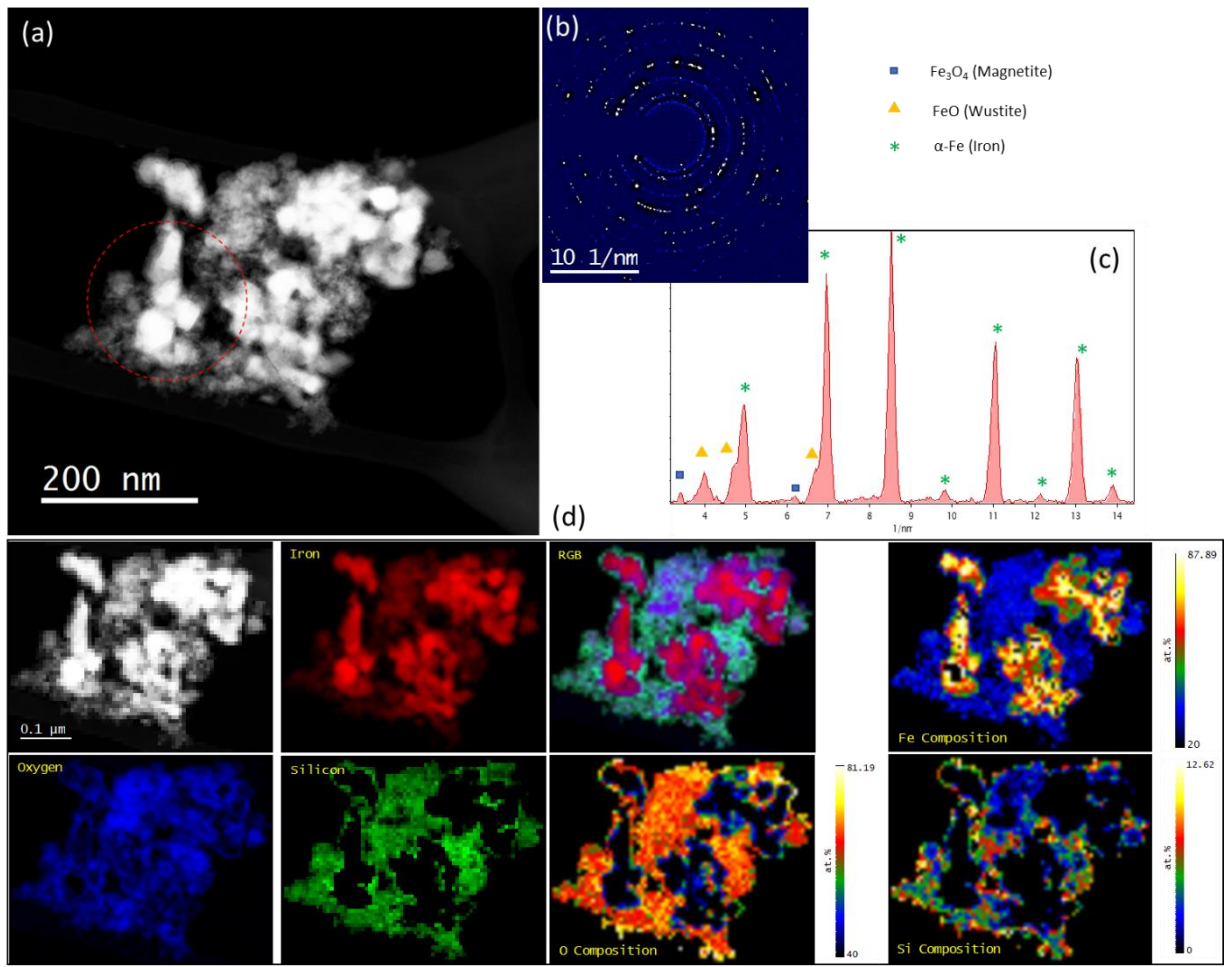

*Figure 8: (a) ADF STEM image of a region in a silica promoted 100/10 Fe/SiO$_2$ after ex situ reduction at 400°C in hydrogen for 20 hours. (b) Filtered SAED pattern obtained from the red dotted circle in (a). (c) Electron diffraction spectrum generated by rotational averaging of (b). (d) Elemental distribution and compositional maps of Fe, O and Si obtained from the region. Added is an RGB false colour overlay of distribution maps.*

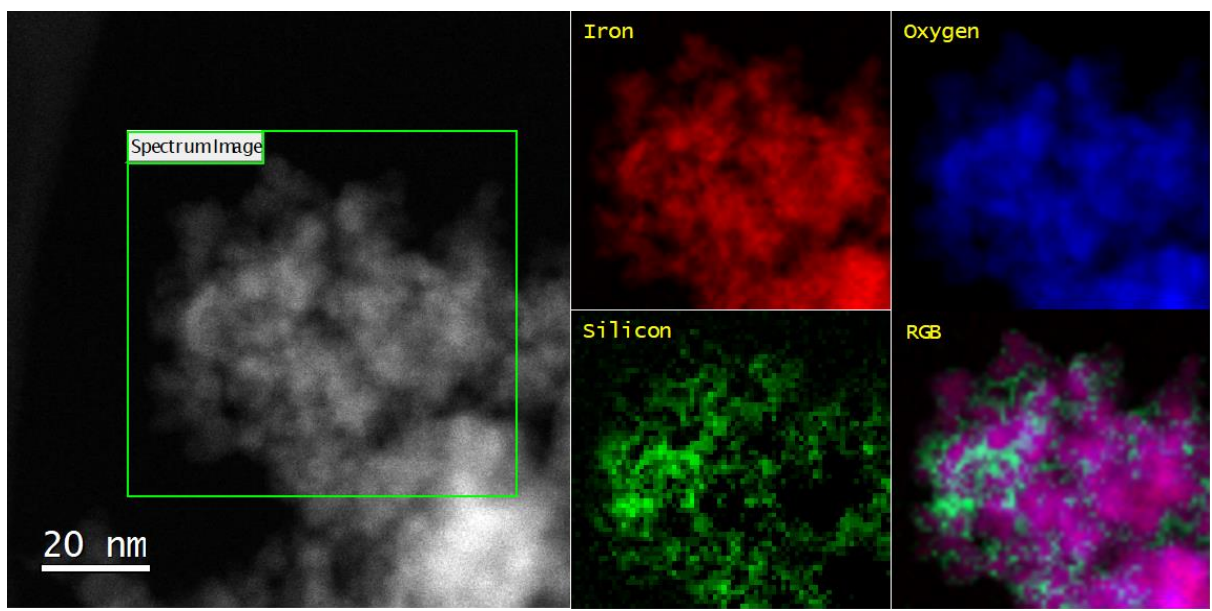

*Figure 9: Elemental distribution maps of iron, oxygen and silicon in a Si-rich zone of a silica promoted 100/10 Fe/SiO2 after* ex situ *reduction at 400°C in hydrogen for 20 hours. Included is a false colour overlay (RGB) of the signals.*

*3.4 Insights into Reduction Mechanism*

From the findings presented above, the reduction mechanism of the silica-promoted, co-precipitated iron(III) oxides to α-Fe and iron(II) silicate may be elucidated. It shows the first direct observations of the structure of such nanoscale materials during the reduction process and confirm the bulk observations made using Mössbauer spectroscopy [10] [8] [11].

Initially, the silica promoted iron oxide exists as a poorly crystalline, silicon-containing iron oxide: an amorphous Fe-O-Si bonded network. On reduction of the $Fe^{3+}$ cations to $Fe^{2+}$, segregation of Fe and Si are initiated with oxygen lost from the Fe-O-Si network as water. The segregation is driven by $Si^{4+}$ cations preferentially substituting within $Fe^{3+}$ sites. The initial formation of crystalline iron oxide phases (predominantly magnetite) occurs within areas of enriched Fe. Sintering of these zones occurs at elevated temperatures with a portion of magnetite transforming into wüstite with additional reduction. Simultaneously, further reduction of $Fe^{3+}$ zones to $Fe^{2+}$ is accompanied by increasing segregation and concentration of Si towards $Fe^{3+}$ zones. The eventual reduction of $Fe^{3+}$ cations to $Fe^{2+}$ cations to $Fe^0$ is thus progressively inhibited at this stage for promoted samples since enhanced segregation and concentration of Si in unreduced areas lead to the formation of distributed nanoscale silica networks among nanocrystalline iron oxide phases (Figure 9). This leaves a silica-rich Fe-O-Si network that is attributed to the iron(II) silicate species that has been observed in previous work [10] [11]. Reduction of magnetite and wüstite to α-Fe occurs at higher temperatures of reduction with the diffusion of isolated zero-valent iron atoms observed. The zero-valent iron atoms agglomerate with time and lead to the formation of distributed metallic Fe nanoparticles. However, a certain portion of wüstite remains difficult to reduce since the formation of a protective silica-containing layer covering remaining iron oxide regions occurs (Figure 9). With sufficient energy (temperature) and time, the remaining amorphous Fe-O-Si regions do form crystalline fayalite, as was observed in the *in situ* TEM analysis after reduction in hydrogen at high temperatures.

## 4. Conclusions

The reduction of unpromoted and silica-promoted iron oxides (100/10 $Fe/SiO_2$) was studied using *in situ* XRD and *in situ* and *ex situ* aberration-corrected TEM. In the absence of silica, hematite (unpromoted) reduces to α-Fe in two stages via magnetite. The 10 wt% silica promotion of the catalyst (100/10 $Fe/SiO_2$) leads to a reduction of the silicon-containing amorphous iron oxide to proceed via a 3-step pathway to α-Fe, with both magnetite and wüstite as stable intermediate phases.The reduction mechanism is discussed, as well as the origin of the iron silicate phase which has been observed historically by bulk analysis techniques. The fundamental insights given here will prove useful in future studies investigating the design of improved Fe-LTFT catalyst materials which could include additional promoters such as potassium and copper.

**Author Information**

*Corresponding author: jaco.olivier@mandela.ac.za

Centre for HRTEM, Nelson Mandela University

Port Elizabeth, South Africa

+27 41 504 4297

**Acknowledgements**

MJC, EJO, JHN and HEdP gratefully acknowledge the DST, NRF and Sasol for funding. SJH acknowledges funding from the engineering and physical sciences research council (UK) EPSRC (grants EP/M010619/1, EP/S021531/1, EP/P009050/1) and the European Commission H2020 ERC Starter grant EvoluTEM (715502).

**Data Availability**

The raw/processed data required to reproduce these findings cannot be shared at this time as the data also forms part of an ongoing study.

**References**

[1] F. Fischer and H. Tropsch, "The synthesis of petroleum at atmospheric pressures from gasification products of coal," *Brennst. Chemie,* vol. 17, no. 3, p. 97, 1926.

[2] M. E. Dry, "The Fischer-Tropsch process: 1950-2000," *Catal. Today,* vol. 71, no. 3-4, pp. 227-241, 2002.

[3] H. Suo, S. Wang, C. Zhang, J. Xu, B. wu, Y. Yang, H. Xiang and Y.-W. Li, "Chemical and structural effects of silica in iron-based Fischer-Tropsch synthesis catalysts," *J. Catal.,* vol. 286, pp. 111-123, 2012.

[4] H. N. Pham and A. K. Datye, "The synthesis of attrition resistant slurry phase iron Fischer-Tropsch catalysts," *Catalysis Today,* vol. 58, pp. 233-240, 2000.

[5] D. B. Bukur, X. Lang, D. Mukesh, W. H. Zimmerman, M. P. Rosynek and C. Li, "Binder/support effects on the activity and selectivity of iron catalysts in the Fischer-Tropsch synthesis," *Ind. Eng.Chem. Res.,* vol. 29, no. 8, pp. 1588-1599, 1990.

[6] K. Jothimurugesan, J. J. Spivey, S. K. Gangwal and J. G. Goodwin, "Effect of silica on iron-based Fischer-Tropsch catalysts," *Stud. Surf. Sci. Catal.,* vol. 119, pp. 215-220, 1998.

[7] C. R. F. Lund and J. A. Dumesic, "Strong oxide-oxide interactions in silica-supported magnetite catalysts. 1. X-ray diffractions and Mossbauer spectroscopy evidence for interaction," *J. Phys. Chem ,* vol. 85, no. 21, pp. 3175-3180, 1981.

[8] A. F. H. Wielers, A. J. H. M. Kock, C. E. C. A. Hop, J. W. Gues and A. M. van der Kraan, "The reduction behaviour of silica-supported and alumina-supported iron catalysts: A Mössbauer and infrared spectroscopic study," *J. Catal.,* vol. 117, no. 1, pp. 1-18, 1989.

[9] D. B. Bukur and C. Sivaaj, "Supported iron catalysts for slurry phase Fischer-Tropsch synthesis," *Appl. Catal. A Gen.,* vol. 231, no. 1-2, pp. 201-214, 2002.

[10] H. Dlamini, T. Motjope, G. Joorst, G. ter Stege and M. Mdleleni, "Changes in physico-chemical properties of iron-based Fischer-Tropsch catalyst induced by SiO2 addition," *Catalysis Letters,* vol. 78, no. 1-4, pp. 201-207, 2002.

[11] C.-H. Zhang, H.-J. Wan, Y. Yang, H.-W. Xiang and Y.-W. Li, "Study on the iron-silica interaction of a co-precipitated Fe/SIO2 Fischer-Tropsch synthesis catalyst.," *Catalysis Communications,* vol. 7, no. 9, pp. 733-738, 2006.

[12] C. R. F. Lund and J. A. Dumesic, "Strong oxide-oxide interactions in silica-supported magnetite catalysts. 2. The core/shell nature of the interaction," *J. Phys. Chem.,* vol. 86, no. 1, pp. 130-135, 1982.

[13] S. Yuen, Y. Chen, J. E. Kubsh and J. A. Dumesic, "Metal oxide-support interactions in silica-supported iron oxide catalsts probed by nitric oxide adsorption," *J. Phys. Chem.,* vol. 86, no. 15, pp. 3022-3032, 1982.

[14] C. R. F. Lund and J. A. Dumesic, "Strong oxide-oxide interactions in silica-supported magnetite catalysts: IV. Catalytic consequences of the interaction in water-gas shift," *Journal of Catalysis,* vol. 76, no. 1, pp. 93-100, 1982.

[15] E. de Smit, I. Swart, J. F. Creemer, G. H. Hoveling, M. K. Gilles, T. Tyliszczak, P. J. Kooyman, H. W. Zandbergen, C. Morin, B. M. Weckhuysen and F. M. F. de Groot, "Nanoscale chemical imaging of a working catalyst by scanning transmission X-ray microscopy," *Nature,* vol. 456, no. 7219, pp. 222-225, 2008.

[16] H. Y. Lin, Y. W. Chen and C. Li, "The mechanism of reduction of iron oxide by hydrogen," *Thermochim. Acta,* vol. 400, no. 1-2, pp. 61-67, 2003.

[17] J. Zieliski, I. Zglinicka, L. Znak and Z. Kaszkur, "Reduction of Fe2O3 with hydrogen," *Appl. Ctal. A. Gen,* vol. 447, no. 1, pp. 191-196, 2010.

[18] A. Venugopal and M. S. Scurrell, "Low temperature reductive pre-treatment of Au/Fe2O3 catalysts, TPR/TPO studies and behavious in the water-gas shift reaction," *Appl. Catal. A Gen.,* vol. 258, no. 2, pp. 241-249, 2004.

[19] R. Cornell and U. Schwertmann, The Iron Oxides: Structure, Properties, Reactions, Occurences and Uses, Weinheim: Wiley-VCHGmbH & Co. KGaA, 2003.

[20] W. Jozwiak, E. Kaczmarek, T. Maniecki, W. Ignaczak and W. Maniukiewicz, "Reduction behavior of iron oxides in hydrogen and carbon monoxide atmospheres," *Appl. Catal. A Gen.,* vol. 326, no. 1, pp. 17-27, 2007.

[21] C. Messi, P. Carniti and A. Gervasini, "Kinetics of reduction of supported nanoparticles of iron oxide," *J. Therm. Anal. Calorim.,* vol. 91, no. 1, pp. 93-100, 2008.

[22] G. Munteanu, L. Ilieva and D. Andreeva, "Kinetic paramters obtained from TPR data for a-Fe2O3 and Au/a-Fe2O3 systems," *Thermochim. Acta,* vol. 291, no. 1-2, pp. 171-177, 1997.

[23] A. Pineau, N. Kanari and I. Gaballah, "Kinetics of reduction of iron oxides by H2. Part I: Low temperature reduction of hematite," *Thermochim. Acta,* vol. 447, no. 1, pp. 89-100, 2006.

[24] S. Stølen, R. Glöckner and F. Grønvold, "Nearly stoichiometric iron monoxide formed as a metastable intermediate in a two-stage disproportionation of quenched wüstite. Thermodynamic and kinetic aspcets," *Thermochimica Acta,* vol. 256, pp. 91-106, 1995.

[25] F. V. Chukhov, B. B. Zvyagin, A. I. Gorshkov, L. P. Yermilova and V. V. Balashova, "Ferrihydrite," *Int. Geol. Rev.,* vol. 16, no. 10, pp. 1131-1143, 1974.

[26] V. A. Drits, B. A. Sakharov, A. L. Salyn and A. Manceau, "Structural model for ferrihydrite," *Clay Miner.,* vol. 28, no. 2, pp. 185-207, 1993.

[27] F. M. Michel, L. Ehm, S. M. Antao, P. L. Lee, P. J. Chupas, G. Liu, D. R. Strongin, M. A. A. Schoonen, B. L. Phillips and J. B. Parise, "The structure of ferrihydrite, a nanocrystalline material," *Science,* vol. 316, no. 5832, pp. 1726-1729, 2007.

[28] C. Masina, J. Neethling, E. Olivier, E. Ferg, S. Manzini, L. Lodya, P. Mohlala and M. Ngobeni, "Mechanism of reduction in hydrogen atmosphere and thermal transformation of synthetic ferrihydrite nanoparticles," *Thermochim. Acta,* vol. 599, pp. 73-83, 2015.

[29] Gatan Inc., "Company page of Gatan Inc., the source of DigitalMicrograph. You can request the free license from this page and download the latest version," [Online]. Available: http://www.gatan.com/.

[30] C. Gammer, C. Mangler, C. Rentenberger and H. Karnthaler, "Quantitative local profile analysis of nanomaterials by electron diffraction," *Scripta Materialia,* vol. 63, no. 3, pp. 312-315, 2010.

[31] Z. Cvejic, S. Rakic, A. Kremenovic, B. Antic, C. Jovalekic and P. Colomban, "Nanosize ferrites obtained from ball milling: crystal structure, cation distribution, size-strain analysis and Raman investigations," *Solid State Science,* vol. 8, no. 8, pp. 908-915, 2006.

[32] H. Fjellvag, F. Gronvold, S. Stolen and B. Hauback, "On the crystallographic and magnetic structures of nearly stoichiometric iron monoxide," *Journal of Solid State Chemistry,* vol. 124, no. 1, pp. 52-57, 1996.

[33] J. R. Smyth, "High temperature crystal chemistry of fayalite," *American Minerologist,* vol. 60, no. 2, pp. 1092-1097, 1975.